\documentclass[10pt,letterpaper]{article}
\usepackage[top=0.85in,left=2.75in,footskip=0.75in]{geometry}
\usepackage{graphicx,amsmath,amssymb,wasysym,verbatim,color}
\newcommand{\be}{\begin{equation}}
\newcommand{\ee}{\end{equation}}
\usepackage{changepage}
\usepackage[utf8x]{inputenc}
\usepackage{textcomp,marvosym}
\usepackage{cite}
\usepackage{nameref,hyperref}
\usepackage[right]{lineno}
\usepackage{microtype}
\DisableLigatures[f]{encoding = *, family = * }
\usepackage[table]{xcolor}
\usepackage{array}
\newcolumntype{+}{!{\vrule width 2pt}}

\newlength\savedwidth

\usepackage[aboveskip=1pt,labelfont=bf,labelsep=period,justification=raggedright,singlelinecheck=off]{caption}

\makeatletter
\renewcommand{\@biblabel}[1]{\quad#1.}
\makeatother

\date{}

\usepackage{lastpage,fancyhdr,graphicx}
\usepackage{epstopdf}
\fancyheadoffset[L]{2.25in}
\fancyfootoffset[L]{2.25in}

\begin{document}
\vspace*{0.2in}

\begin{flushleft}
{\Large
\textbf\newline{Correlated network of networks enhances robustness against catastrophic failures}
}
\newline
\\
Byungjoon Min\textsuperscript{1,2*},
Muhua Zheng\textsuperscript{2,3}
\\
\bigskip
\textbf{1} Department of Physics, Chungbuk National University, Cheongju, Chungbuk 28644, Korea
\\
\textbf{2} Levich Institute and Physics Department, City College of New York, New York, New York 10031, USA
\\
\textbf{3} Department of Physics, East China Normal University, Shanghai, 200062, China
\\
\bigskip


* min.byungjoon@gmail.com 

\end{flushleft}
\section*{Abstract}
Networks in nature rarely function in isolation but instead interact with one another 
with a form of a network of networks (NoN). A network of networks with interdependency 
between distinct networks contains instability of abrupt collapse related to 
the global rule of activation. 
As a remedy of the collapse instability,
here we investigate a model of correlated NoN. 
We find that the collapse instability can be removed when hubs provide the majority 
of interconnections and interconnections are convergent between hubs.
Thus, our study identifies a stable structure of correlated NoN against catastrophic 
failures. Our result further suggests a plausible way to enhance network robustness
by manipulating connection patterns, along with other methods such as controlling 
the state of node based on a local rule.



\section*{Introduction}

Real-world complex systems ranging from critical infrastructure \cite{little,rosato,buldyrev} 
and transportation networks \cite{domenico,parshani} to living organisms 
\cite{reis,vidal,white} are rarely formed by an isolated network but 
by a network of networks (NoN) \cite{buldyrev,reis,Souza,non_model,kevin,
brummitt,li,Parshani2010,Parshani2011,Son2010,Baxter2012,Dong2012,Son2012,
Zhou2012,Valdez2013,Zhao2013,via,Radicchi2015,zwang2,zwang,wang,wang2,zhang,liu}. 
For instance, different kinds of critical infrastructures such as a power grid 
and the Internet are coupled and interact with one another \cite{little,rosato}.
In addition, many living systems including brain networks \cite{reis,dosenbach}
and cellular networks \cite{vidal} consist of different modules 
strongly connected and interconnections between them.

Several models of a system of networks have been proposed with the role of 
interconnections that are links across different 
networks~\cite{buldyrev,Souza,non_model}. Models of NoN may 
fall into three classes according to the functionality of interconnections: 
Modular NoN (M-NoN), Catastrophic NoN (C-NoN), and Robust NoN (R-NoN).
A primitive model of NoN is Modular NoN
in which intraconnections within a network and interconnections between 
different networks have no difference in function \cite{Souza}. 
Since nodes connected by an interconnection do not control each other,
this model corresponds to a single modular network 
with a different density of intraconnections and interconnections.

However, considering distinct nature of intraconnections and interconnections in NoN, 
a different role for different types of connections may be required.
For example, when different networks function interdependently,
interconnections should not play the same role as intraconnection
but control the state of a connected node in the other networks \cite{buldyrev,non_model}.
And, the state of a node in C-NoN model is determined by the global
characteristics of a network \cite{buldyrev,gao}. 
To be specific, a node can be active only if any interconnected nodes in 
different networks belong to the global giant component. 
Such global rule results in an extreme instability of a system of networks 
since a small perturbation can trigger catastrophic collapse.

In order to resolve the conflict between the extreme fragility 
and robust systems of networks observed widely in reality 
such as the brain, R-NoN model in which the state of a node is controlled by local 
property of interconnected nodes have been proposed \cite{non_model,kevin}. 
For R-NoN, nodes connected by an interconnection still control each other.
But, a node in R-NoN model can be active even though interconnected nodes in 
a different network do not belong to the global giant component. 
With this modification, R-NoN model becomes robust but still maintains the 
functionality across different networks.

Beside R-NoN, it is of interest how to produce a more robust C-NoN system
because there are some examples to follow the global rule such as a power grid.
Catastrophic NoN model involves vulnerability related to the global
rule leading to the potential danger of abrupt collapse.
Here, we investigate a modified model taking into account a correlation
in the connectivity patterns of NoN as a remedy of the collapse instabilities.
So far, the majority of research about networks of networks have studied 
NoN with uncorrelated and one-to-one interconnections \cite{buldyrev,gao}. 
In contrast, a system of coupled networks in reality are composed with 
one-to-many interconnections and a degree-degree correlation between nodes 
in distinct networks \cite{parshani,reis,nicosia,szell,bmin}.
For instance, for the case of the brain networks, non-trivial patterns
of connections have been reported for 
resting state and in task \cite{reis}.
Correlated coupling was also observed in several different types 
of complex systems such as transportation networks \cite{nicosia},
social networks \cite{szell}, and critical infrastructure networks \cite{rosato,bmin}.

In this study we find that the collapse instabilities in C-NoN 
can be removed, and the model becomes stable by introducing correlated NoN.
Specifically, we investigate the effect of a degree-degree correlation 
on network robustness under random removal of nodes
by extending a previous analysis \cite{reis}.
We find that when hubs are major source of outgoing links
and the interconnections are convergent between hubs,
NoN becomes stable to function properly.
Our study provides an optimal design of correlated NoN against 
an external perturbation and a possible reason for stable 
functioning of correlated NoN in reality.

\section*{Model and Theory}

We consider a network of networks composed of two 
networks, $A$ and $B$, with interconnections between 
the networks, for the sake of simplicity.
Each node in NoN can have two different types of links,
inlinks and outlinks. Inlink refers connections inside 
the same network while outlink is connections 
between nodes in different networks.

Here, we examine two different modes of interactions of 
out-links \cite{reis}: Catastrophic NoN and Modular NoN. 
C-NoN represents that a node in network $A$ 
operates properly only when one of the reciprocal nodes in network $B$
connected by outlinks also functions properly.
Thus, a node in network $A$ cannot be active when 
it does not belong to the giant component on network $A$ or it 
loses all connectivity to network $B$.
On the other hand, for M-NoN, a node in 
network $A$ can be active if it belongs to the giant component 
through either inlink or outlink.
Thus, even though a node in network $A$ is completely 
decoupled from network $B$, such node can be active as long as it 
still belongs to the giant component of $A$.
Therefore, for M-NoN mode of interactions, there is no 
cascading failure after the initial removal of nodes.

\begin{figure}
\begin{center}
\includegraphics[width=\linewidth]{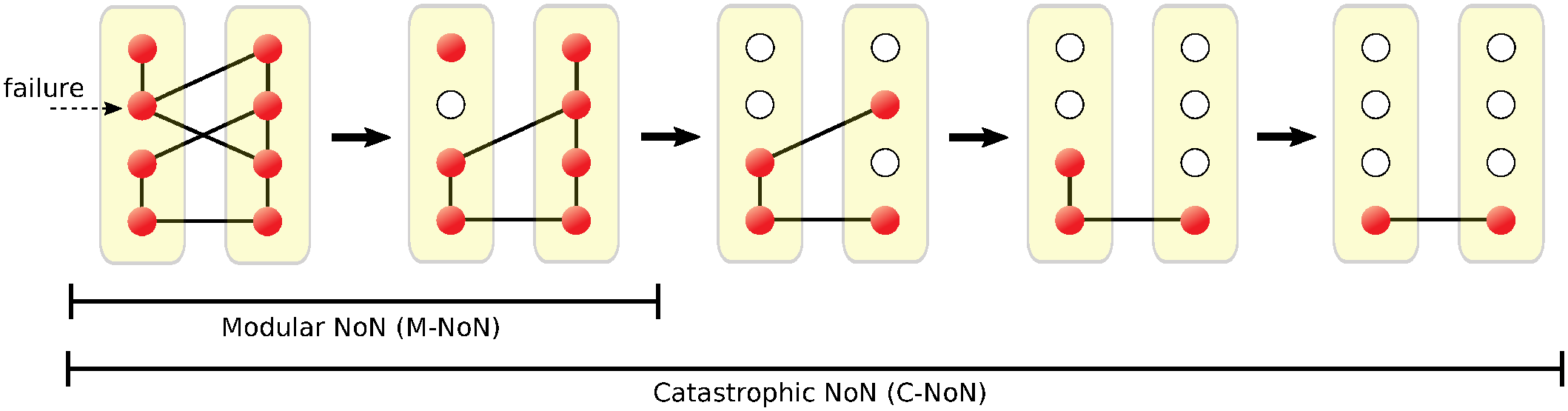}
\caption{
An example of C-NoN and M-NoN. In this example, initially a single node is 
removed by an external perturbation. For M-NoN, this node and all of 
its links are removed. For C-NoN, we further remove nodes and 
their connections if they do not have any interconnections.
These removing processes proceed iteratively until there are no more 
nodes to be removed.
}
\label{fig0}
\end{center}
\end{figure}

An example of C-NoN and M-NoN is depicted in Fig.~\ref{fig0}.
For M-NoN model, a fraction of nodes are targeted to be removed. 
Then all targeted nodes and their connections are removed from
the original network. Finally we identify the largest connected 
component linked by either intraconnections or interconnections.
For C-NoN model, after the removal of initially targeted nodes,
we further remove nodes that do not have any interconnections.
In addition, we remove all nodes that do not belong 
to the largest connected component. So, we remove 
iteratively nodes that do not belong to the giant component
or do not have any interconnections. These removal processes lead to 
cascading failure.

In order to assess the robustness of a system against 
random removal, we measure the size of giant component after initial node removal.
We also identify the percolation threshold $p_c$ 
at which the giant component disappears, to measure stability of NoN.
NoN with low threshold corresponds to stable structures
because many nodes need to be removed to break it down,
whereas high percolation threshold represents vulnerable structures.

\subsection*{Catastrophic network of networks}
In this section, we introduce a theory for C-NoN 
mode of interactions to find the size of giant component
and percolation threshold.
Initially, all nodes in 
both networks $A$ and $B$ are active.
A fraction $p_A$ and $p_B$ of nodes randomly chosen are 
removed from the networks $A$ and $B$, respectively.
Then, a node is active only if it belongs to the giant 
component in its network via in-links and at the same time 
connects to the giant component on the other network via 
one of its out-links.
Nodes that do not satisfy the survival condition are 
removed from NoN iteratively.
Note that nodes that do not have any out-links at the 
beginning can be active as long as they remain to connect 
with the giant component via in-links.

To obtain the percolation threshold $p_c$, we introduce 
a joint degree distributions of indegree and outdegree as $P(\vec{k})$ 
where $\vec{k}=(k_{in}^A,k_{in}^B,k_{out}^A,k_{out}^B)$.
We also introduce a conditional degree distribution for 
a pair of connected nodes in different networks to take into account 
a degree-degree correlation,
$P_{AB}(k_{in}^A|k_{in}^B)$ and $P_{BA}(k_{in}^B|k_{in}^A)$.
Next, we develop a theoretical framework for the robustness of NoN on a locally 
tree-like structure with an arbitrary joint degree distribution and 
a conditional degree distribution \cite{reis}.

We define $u_{A}$ and $u_{B}$ respectively as the probability that a node 
in networks $A$ and $B$ reached by a randomly chosen in-link does 
not belong to a mutually connected giant component. 
$u_A$ and $u_B$ can be expressed by the following self-consistency equation
\begin{eqnarray}
1-u_i=&p_i& \Bigg[\sum_{\vec{k}} \frac{k_{in}^i P(\vec{k})}{\langle k_{in}^i \rangle} (1-u_{i}^{k_{in}^i-1})(\delta_{k_{out}^i,0}+1-w_{k_{in}^i}^{k_{out}^i}) \Bigg],
\end{eqnarray}
where $i\in \{A,B\}$ and $\delta_{i,j}$ is the Kronecker delta.
Here, $w_{k_{in}^i}$ is the probability that a node reached by a randomly 
chosen out-link from a node in network $i$ with indegree $k_{in}^i$ 
does not belong to the giant component of the opposite network.
The first term $(1-u_{i}^{k_{in}^i-1})$ represents the probability 
that a node with $k_{in}^i$ belongs to the giant component in 
network $i$, and the second term represents that the probability 
that a node with $k_{in}^i$ connects with the giant component of 
the opposite network through an outlink.
By the term $\delta_{k_{out}^i,0}$ in Eq.~(1),
a node without out-links ($k_{out}^i=0$) can be treated 
differently with other nodes ($k_{out}^i \ne 0$).
Then, the probability $w_{k_{in}^i}$ can be expressed as
\begin{eqnarray}
1-w_{k_{in}^i}=p_i \left[ 1- \sum_{k_{in}^i} P(k_{in}^j|k_{in}^i) u_{j}^{k_{in}^j} \right].
\end{eqnarray}
Obtaining $u_{i}$ and $w_{k_{in}^i}$ by solving these equations, 
the size $G_i$ of the mutually connected giant component of 
C-NoN is given by
\begin{equation}
G_i=p_i \left[ \sum_{\vec{k}} P(\vec{k}) (1-u_i^{k^i_{in}})(\delta_{k^i_{out},0}+1-w_{k_{in}^i}^{k^i_{out}}) \right].
\end{equation}

\subsection*{Modular network of networks}
For M-NoN, a node can survive if it belongs 
to the giant component a whole network. 
Given degree distributions, the probability $\nu_i$ that 
a node reached by a randomly chosen inlink of network $i$ 
does not belong to the giant component of M-NoN 
is given by 
\begin{eqnarray}
1-\nu_i=p_i \left[ \sum_{\vec{k}} \frac{k^i_{in} P(\vec{k})}{\langle k^i_{in} \rangle} (1-\nu_i^{k^i_{in}-1} \mu_{k_{in}^i}^{k_{out}^i}) \right].
\end{eqnarray}
Here, $\mu_{k_{in}^i}$ is the probability that a node reached 
by a randomly chosen outlink from a node in network $i$ with 
indegree $k_{in}^i$ does not belong to the giant component of 
the opposite network.
And, the probability $\mu_{k_{in}^i}$ can be obtained by following,
\begin{eqnarray}
1-\mu_{k_{in}^i}=p_i \left[ 1- \sum_{k_{in}^i} P(k_{in}^j|k_{in}^i) \nu_{j}^{k_{in}^j} \right].
\end{eqnarray}
For M-NoN, a node in network $i$ can survive 
if it belongs to the giant component in network $i$ or 
the giant component in a different network by an interconnection.
Once we obtain $\nu_i$ and $\mu_{k_{in}^i}$, 
the size $G_i$ of the giant component of M-NoN is 
\begin{eqnarray}
G_i=p_i \left[ \sum_{\vec{k}} P(\vec{k}) (1-\nu_i^{k^i_{in}} \mu_{k_{in}^i}^{k_{out}^i}) \right].
\end{eqnarray}

\subsection*{Correlation in network of networks}

\begin{figure}
\begin{center}
\includegraphics[width=0.65\linewidth]{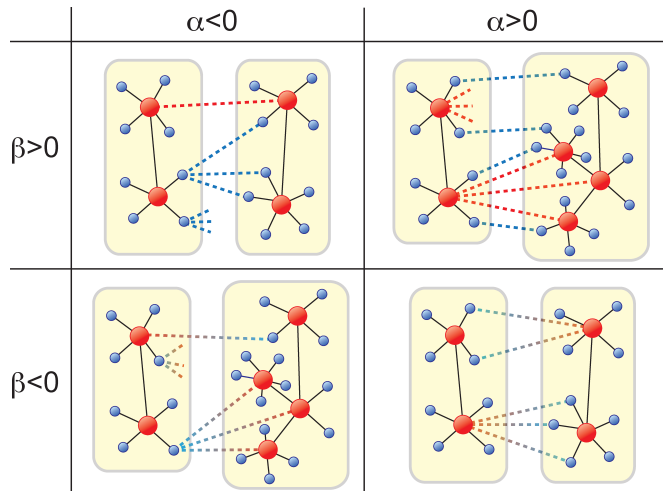}
\caption{
Diagram of a correlated network of networks 
according to parameters $\alpha$ and $\beta$.
Hubs (red nodes) and non-hubs nodes (blue nodes) can have 
inlinks (solid lines) and outlinks (dotted lines).
When $\alpha>0$, hubs are more likely to have many outlinks
whereas when $\alpha<0$, non-hub nodes are more likely to have outlinks.
When $\beta>0$, hubs prefer to connect with other hubs in a different network
but when $\beta<0$, hubs in one network prefer to connect to non-hub nodes in a different network. 
}
\label{fig1}
\end{center}
\end{figure}

In real-world complex systems, NoN are not made randomly but with a certain 
degree-degree correlation. Correlated coupling is observed in several 
different kinds of complex systems such as transportation networks \cite{nicosia},
social networks \cite{szell}, and critical infrastructure 
networks \cite{rosato,bmin}, and 
crucial for structural and dynamical properties of networks \cite{vespignani,gallos08,radicchi}.
For instance, functional brain networks of the human
show a peculiar correlation pattern \cite{reis}.
In this paper, we consider a degree-degree correlation using
two scaling parameters, $\alpha$ and $\beta$ (Fig.~\ref{fig1}) as observed in 
functional networks of the human brain \cite{reis}.
The parameter $\alpha$ is defined as 
\begin{eqnarray}
k_{out} \sim k_{in}^{\alpha}.
\end{eqnarray}
Thus, for $\alpha>0$ hubs of each network also have many outlinks, 
whereas for $\alpha<0$ nodes with low degree have many outlinks (Fig.~\ref{fig1}).
The other parameter $\beta$ is defined as
\begin{eqnarray}
k_{in}^{nn} \sim k_{in}^{\beta},
\end{eqnarray}
where $k_{in}^{nn}$ is the average indegree of the nearest neighbors 
in the other network.
Therefore, $\beta$ quantifies indegree-indegree correlation between 
two connected nodes by interconnections.
For $\beta>0$, hubs connect with other hubs in the different network.
Instead for $\beta<0$, hubs in a network connect with nodes with 
less degree in the other network (Fig.~\ref{fig1}).
Note that uncorrelated NoN corresponds to $\alpha=0$ and $\beta=0$.

\section*{Results}
\subsection*{Effect of the density of out-links}
We first examine the robustness of NoN by changing the 
density of links in order to check the effect of outlinks.
As an instructive example, we consider a coupled Erd\"{o}s-R\'enyi 
(ER) network. For ER NoN with no degree correlation, a joint degree distribution 
can be factorized as $P(\vec{k})=P_{in}(\vec{k}_{in}) P_{out}(\vec{k}_{out})$
and a conditional degree distribution can be simply expressed as 
$P(k_{in}^j|k_{in}^i)=P_{in}(k_{in})$.
We assume that two networks have the same average in-degree,
$\langle k_{in}^A \rangle = \langle k_{in}^B \rangle =\langle k_{in}\rangle$,
and the fraction of removed nodes are the same for both networks, 
$p_A=p_B=p$. Then, Eqs. (1) and (2) can be simply reduced into a single equation:
\begin{eqnarray}
u=1-p \left[1-e^{\langle k_{in} \rangle (u-1)} \right]  \left[e^{-\langle k_{out} \rangle} +1 -e^{p \langle k_{out} \rangle (e^{\langle k_{in} \rangle (u-1)}-1)} \right].
\end{eqnarray}
where $\langle k_{out} \rangle$ is the average outdegree.
Once we define the function
\begin{eqnarray}
f(u)=u-1+p \left[1-e^{ \langle k_{in} \rangle (u-1)} \right] \left[e^{-\langle k_{out} \rangle } +1 -e^{p\langle k_{out} \rangle (e^{\langle k_{in} \rangle (u-1)}-1)} \right],
\end{eqnarray}
one can obtain the percolation threshold $p_c$ by imposing the conditions $f(u)=f'(u)=0$.
In addition, a tricritical line $(\langle k_{in} \rangle,\langle k_{out} \rangle,p)$ between
continuous and discontinuous transitions can be computed by the conditions $f(u)=f'(u)=f''(u)=0$.

For M-NoN, the self-consistency equation is similarly  given by
\begin{eqnarray}
1-\nu &=& p \left[1-e^{\langle k_{in} \rangle(\nu-1)} e^{\langle k_{out} \rangle p(e^{ \langle k_{in} \rangle (\nu-1)}-1)} \right]. 
\end{eqnarray}
Then, one can obtain the percolation threshold with the 
conditions $g(\nu)=g'(\nu)=0$, if we define 
\begin{eqnarray}
g(\nu) = \nu -1 + p \left[1-e^{\langle k_{in} \rangle(\nu-1)} e^{ \langle k_{out} \rangle p(e^{\langle k_{in} \rangle(\nu-1)}-1)} \right].
\end{eqnarray}
Note that the percolation transition of M-NoN is 
always second-order and hence a tricritical point does not exist.

\begin{figure}
\begin{center}
\includegraphics[width=0.8\linewidth]{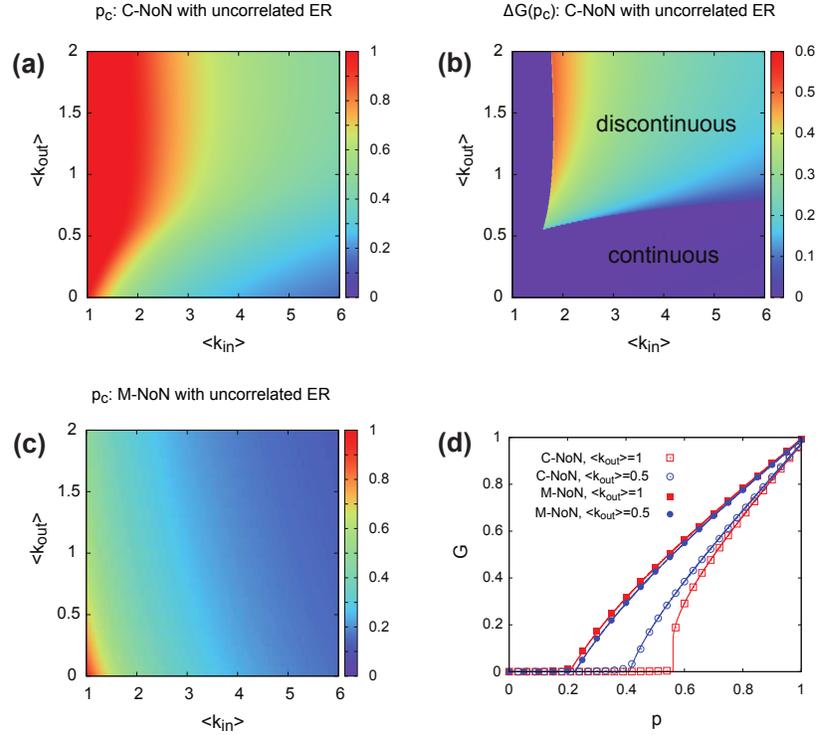}
\caption{
(a) Percolation threshold $p_c$ of C-NoN for two coupled ER networks 
with no correlation predicted by theory.
For high $\langle k_{out} \rangle$ and low $\langle k_{in} \rangle$, 
NoN is stable to maintain mutual connectivity under the random removal of nodes.
(b) The size of jump at the percolation threshold of C-NoN. 
The size of jump shows undergoes a second-order phase transition 
for small $\langle k_{out} \rangle$, but
the transition becomes discontinuous as $\langle k_{out} \rangle$ increases.
(c) Percolation threshold $p_c$ of M-NoN for 
ER NoN with no degree correlation. NoN becomes more stable with increasing
either $\langle k_{in} \rangle$ or $\langle k_{out} \rangle$.
(d) The size of giant component for both C-NoN (open symbol) and 
M-NoN (filled symbol) modes of interactions as a function $p$ of 
a fraction of removed nodes.
Analytic calculation (line) and numerical simulation (symbols) are 
shown together.
}
\label{fig2}
\end{center}
\end{figure}

Increasing the density of out-links, NoN with catastrophic interactions 
becomes getting vulnerable as depicted in Fig.~\ref{fig2}(a).
In addition, the transition between percolating and non percolating phases
becomes discontinuous above a tricritical line and the size of discontinuous
jump at the transition increases with 
increasing $\langle k_{out} \rangle$ [Fig.~\ref{fig2}(b)]. 
For C-NoN, outlinks force interconnected systems 
to be more vulnerable and prone to abrupt collapse due to cascading failure.
On the other hand, inlinks preserve the connectivity and produce more 
robust structures.
In conclusion, NoN with high $\langle k_{in} \rangle$ and low 
$\langle k_{out} \rangle$ shows a stable structure 
for C-NoN.

For M-NoN, however, outlinks play the opposite role.
High density of outlinks enhances network robustness
by adding a potential detour for connectivity [Fig.~\ref{fig2}(c)]. 
Outlinks contribute to maintain the robustness of networks for M-NoN
but they can cause the opposite effect for C-NoN. 
Thus, the optimal design of interconnections between networks
is called for maintaining stable functioning 
for both M-NoN and C-NoN.

\subsection*{Generating correlated networks of networks}

In order to examine the effect of a degree-degree correlation,
we first construct NoN with a correlation $(\alpha,\beta)$.
We construct a network drawn from an indegree distribution $P_i(k_{in})$,
by following configuration model.
Next, stubs of outgoing links are assigned to each node 
with the probability proportional to $k_{in}^{\alpha}$.
Connecting two nodes in different networks with a relationship 
$k_{in}^{nn} \sim k_{in}^{\beta}$ is non-trivial.
We cannot simply assign a set of connections 
for outlinks from a joint distribution $P(\vec{k})$ since 
such a set almost certainly fails to satisfy the topological constraint
because of the reciprocal relation between
$k_{in}^{nn} \sim k_{in}^{\beta}$ and 
$k_{in}^{nn} \sim k_{in}^{1/\beta}$, 
except for $\beta=0$ and $\beta=1$.

Instead, we use the following way as in \cite{reis} to construct NoN
with a degree-degree correlation $\beta$.
We choose randomly node $i$ in network $A$ if it has available outlinks.
Next, we connect node $i$ with node $j$ with degree $k_{in}^B$ 
in network $B$ with the probability that follows a Poisson distribution $P(k_{in}^j)$
with a mean value $\lambda=\langle C_{\beta} k_{in}^\beta \rangle$ where
$C_{\beta}=k_{max}^{(1-\beta)/2}$.
This processes repeat until there are no more out-links left.
This algorithm cannot make NoN with exactly corresponding $\beta$ for most sets
of $(\alpha,\beta)$, but it can guarantee that numerically generated $\beta_{gen}$ 
increases or decreases in a monotonic manner with changing $\beta$ [Figs.~\ref{fig3}(d) and \ref{fig4}(d)].

\subsection*{Robustness of correlated networks of networks} 

To search robust structures of correlated NoN, 
we generate NoN with the above algorithm and 
obtain joint and conditional degree distributions
from the realized networks with $(\alpha,\beta)$.
Next, we identify the critical fraction $p_c$ of nodes removal 
by imposing the condition $G(p_c)=0$,
showing network robustness with a given correlation.
In order to examine the effect of the correlated structure of NoN,
we calculate $p_c(\alpha,\beta)$ for the both modes of C-NoN 
and M-NoN with ER networks and scale-free (SF) networks.
The small $p_c(\alpha,\beta)$ represents robust structures 
against an external perturbation.

\begin{figure}
\begin{center}
\includegraphics[width=0.8\linewidth]{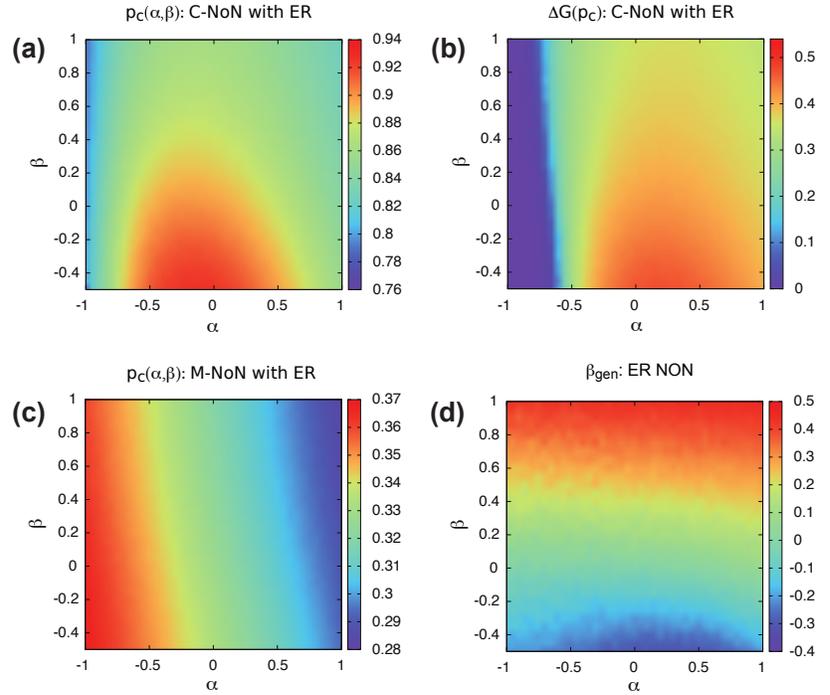}
\caption{
(a) Percolation threshold and (b) size of jump of C-NoN 
in correlated ER NoN with $N=10^4$, $\langle k_{in} \rangle =2$, 
and $\langle k_{out} \rangle =1$ for different $\alpha$ and $\beta$.
When $\alpha \approx -1$ or $\alpha >0.5$ and $\beta>0$,
NoN becomes stable against random failure.
In contrast, when $-0.5<\alpha<0.5$ and $\beta<0$, NoN 
is vulnerable to catastrophic collapse.
(c) percolation threshold of M-NoN with correlated ER NoN
with the same parameters as C-NoN. 
High $\alpha$ and $\beta$ region is robust against random failure for M-NoN.
(d) $\beta_{gen}$ observed from realized networks at a given $(\alpha,\beta)$.
The value $\beta_{gen}$ is obtained by a linear regression.
}
\label{fig3}
\end{center}
\end{figure}

For ER NoN, when $\alpha\approx -1$, low $p_c$ is observed 
regardless of $\beta$, indicating stable NoN [Fig.~\ref{fig3}(a)].
In this region, hubs are isolated in a single network 
and maintain effectively the giant component.
As a result, the extensive size of jump at $p_c$ vanishes [Fig.~\ref{fig3}(b)].
Another stable region is located at $\alpha >0.5$ and $\beta>0$.
High $\alpha$ and $\beta$ guarantees that many hub-hub interconnections,
so that hubs are more likely protected from cascading failure.
When $-0.5<\alpha<0.5$ and $\beta<0$, a system of networks is 
highly vulnerable to catastrophic cascading failure.
With these parameters, hubs connect to nodes with less degree nodes in 
the other network, leading to that hubs can be easily attacked by interdependency. 
For M-NoN, the network robustness enhances with 
increasing $\alpha$ and $\beta$ monotonically [Fig.~\ref{fig3}(c)].
When $\alpha>0$ and $\beta>0$, both inlinks and outlinks converge toward hubs
and the giant component can be preserved with only a few hubs.
Therefore, high $\alpha$ and $\beta$ region
is robust against random failure for M-NoN.

\begin{figure}
\begin{center}
\includegraphics[width=0.8\linewidth]{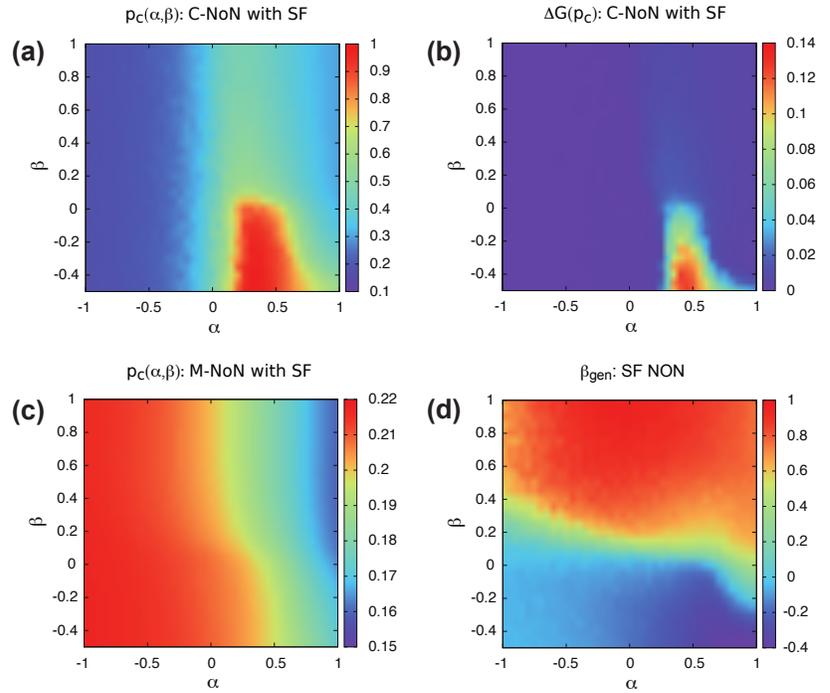}
\caption{
(a) Percolation threshold, (b) size of jump for C-NoN, 
and (c) percolation threshold for M-NoN with two coupled SF networks
with $N=10^4$, $\langle k_{out} \rangle =1$, the degree 
exponent $\gamma=2.5$, and $k_{max}=100$ 
for different $\alpha$ and $\beta$.
High $\alpha$ and $\beta$ region is robust against random failure for 
both C-NoN and M-NoN.
When $\alpha<0$ or $\alpha>0.5$ and $\beta>0$,
NoN becomes stable against random failure.
In contrast, when $-0.5<\alpha<0.5$ and $\beta<0$, NoN 
is vulnerable to catastrophic collapse.
(d) $\beta_{gen}$ obtained by a linear regression from realized 
networks at a given $(\alpha,\beta)$.
}
\label{fig4}
\end{center}
\end{figure}

The impact of the correlation is more clear in SF networks
because of a key role of hubs with an inhomogeneous degree distribution.
When $\alpha<0$, a networked system is stable (low $p_c$) 
because hubs are protected from cascading failure for N-NoN [Fig.~\ref{fig4}(a)].
When $\alpha >0.5$ and $\beta>0$, networks are also stable 
since hubs are more likely active due to a lot of interconnections between them.
However, for intermediate $\alpha$ ($0<\alpha<0.5$) and 
divergent interconnections ($\beta<0$),
hubs are easily exposed to cascading failure since 
they connect to non-hub nodes in the other network.
In this region, C-NoN is fragile to random attack and results in abrupt collapse 
as shown in Fig.~\ref{fig4}(b).
For M-NoN, a coupled SF network is more vulnerable when $\alpha<0$ because hubs 
have only few outlinks as in ER NoN [Fig.~\ref{fig4}(c)].

In conclusion, the degree-degree correlation in NoN allows us to find
a stable structure for functioning of NoN.
When hubs have many interconnections ($\alpha \approx 1$) and hub-hub 
interconnections are abundant ($\beta>0$), NoN can maintain a robust structure 
for both C-NoN and M-NoN. 
And, M-NoN is vulnerable when $\alpha<0$
and C-NoN is at risk of catastrophic collapse when $\beta<0$.

\section*{Discussion}

We study the robustness of a system of networks with degree-degree 
correlations and one-to-many interconnections between distinct networks.
We investigate the effect of degree-degree correlations on the network robustness
with different modes of interconnections. 
For uncorrelated NoN, outlinks reduce the robustness for C-NoN 
while they enhance the robustness for M-NoN. However, taking into account 
the degree correlation, we find stable structures in correlated networks 
of networks for both C-NoN and M-NoN. 
Specifically, when hubs provide most interconnections 
and the interconnections are convergent, networks of networks 
become more robust for both modes of interconnections. 
Our study of correlated NoN can shed light on finding the origin of reliable 
functioning of interconnected networks in reality. In addition, it can provide
an economical method of designing robust multilayered systems such as 
interconnected infrastructures or financial systems.
In addition to correlated NoN, 
robust NoN model which is recently proposed \cite{non_model,kevin}
can be another plausible solution of stable functioning of NoN 
and also allow us to find the core areas in NoN 
\cite{non_model,morone,plosone,cip,epjb,kitsak,pei2,pei1}.


\nolinenumbers

\newpage

\end{document}